\documentclass[aip,apl,reprint]{revtex4-1}
\usepackage{amsfonts}
\usepackage{amsmath}
\usepackage{amssymb}
\usepackage{graphicx}

\usepackage[bookmarks=true,bookmarksopen,bookmarksnumbered,
			colorlinks,linkcolor=blue,citecolor=blue,
            breaklinks=true]{hyperref}

\newcommand{\Nmin}{N_{\text{min}}}
\newcommand{\grep}{\gamma_{\text{rep}}}

\begin{document}

\title{Electron Spin Resonance spectroscopy with femtoliter detection volume} %Title of paper

\author{V.~Ranjan}

\affiliation{Quantronics group, SPEC, CEA, CNRS, Universit\'e Paris-Saclay, CEA Saclay 91191 Gif-sur-Yvette Cedex, France}

\author{S.~Probst}
\affiliation{Quantronics group, SPEC, CEA, CNRS, Universit\'e Paris-Saclay, CEA Saclay 91191 Gif-sur-Yvette Cedex, France}

\author{B.~Albanese}
\affiliation{Quantronics group, SPEC, CEA, CNRS, Universit\'e Paris-Saclay, CEA Saclay 91191 Gif-sur-Yvette Cedex, France}

\author{T.~Schenkel}
\affiliation{Accelerator Technology and Applied Physics Division, Lawrence Berkeley National Laboratory, Berkeley, California 94720, USA}

\author{D.~Vion}
\affiliation{Quantronics group, SPEC, CEA, CNRS, Universit\'e Paris-Saclay, CEA Saclay 91191 Gif-sur-Yvette Cedex, France}

\author{D.~Esteve}
\affiliation{Quantronics group, SPEC, CEA, CNRS, Universit\'e Paris-Saclay, CEA Saclay 91191 Gif-sur-Yvette Cedex, France}

\author{J.~J.~L.~Morton}
\affiliation{London Centre for Nanotechnology, University College London, London WC1H 0AH, United Kingdom}

\author{P.~Bertet}
\email[]{patrice.bertet@cea.fr}
\affiliation{Quantronics group, SPEC, CEA, CNRS, Universit\'e Paris-Saclay, CEA Saclay 91191 Gif-sur-Yvette Cedex, France}

\date{\today}

\begin{abstract}
We report electron spin resonance measurements of donors in silicon at millikelvin temperatures using a superconducting $LC$ planar micro-resonator and a Josephson Parametric Amplifier. The resonator includes a nanowire inductor, defining a femtoliter detection volume. Due to strain in the substrate, the donor resonance lines are heavily broadened. Single-spin to photon coupling strengths up to $\sim 3~\text{kHz}$ are observed. The single shot sensitivity is $120 \pm 24~$spins/Hahn echo, corresponding to $\approx 12 \pm 3$~spins$/\sqrt{\text{Hz}}$ for repeated acquisition.
\end{abstract}

%%07.57.Pt: Submillimeter wave, microwave and radiowave spectrometers; magnetic %resonance spectrometers, auxiliary equipment, and techniques
%76.30.-v: Electron paramagnetic resonance and relaxation (see also 33.35.+r %Electron resonance and relaxation in atomic and molecular physics; 87.80.Lg %Magnetic and paramagnetic resonance in biological physics)
%85.25.-j: Superconducting devices

\maketitle

Electron spin resonance (ESR) spectroscopy is useful for characterizing paramagnetic species and finds applications in a large number of fields. The most widely used detection method is the so-called inductive detection, which relies on the emission of microwave signals by the spins during their Larmor precession into a resonant cavity to which they are magnetically coupled. Conventional inductively-detected ESR spectroscopy suffers from a low spin detection sensitivity, which precludes its use for micron- or nano-scale samples~\cite{schweiger_principles_2001}, motivating research on alternative detection schemes~\cite{wrachtrup_optical_1993,gruber_scanning_1997,rugar_mechanical_1992,rugar_single_2004,chamberlin_high-sensitivity_1979,manassen_direct_1989,baumann_electron_2015,morello_single-shot_2010}. 

Planar micro-resonators~\cite{narkowicz_scaling_2008,artzi_induction-detection_2015} and self-resonant microhelices~\cite{sidabras_extending_2019} have been shown to be promising to push inductive detection to higher sensitivity and lower detection volumes, but the microwave confinement that they enable is ultimately limited by ohmic losses in the metal. This can be overcome by the use of superconducting micro-resonators at low temperatures~\cite{malissa_superconducting_2013,sigillito_fast_2014,bienfait_reaching_2015,eichler_electron_2017}, for which arbitrarily small detection volumes should in principle be achievable while preserving a high resonator quality factor. An additional benefit of using small-mode-volume and high-quality-factor resonators is the enhanced microwave spontaneous emission they cause via the Purcell effect~\cite{purcell_spontaneous_1946,goy_observation_1983,bienfait_controlling_2016,eichler_electron_2017}, which enables to repeat measurements faster and therefore impacts favorably the spin detection sensitivity. A recent experiment detected the ESR signal from an ensemble of donors in silicon at millikelvin temperatures coupled to a superconducting resonator with a sub-pL magnetic mode volume, reaching a spin detection sensitivity of 65 spin/$\sqrt\text{Hz}$ measured by a Hahn-echo sequence~\cite{probst_inductive-detection_2017}. 

Here, we push this effort further with a new resonator geometry incorporating a superconducting nanowire around which the magnetic component of the microwave field is confined, yielding a mode volume as low as $\sim 6$\,fL. Correspondingly, the spin-photon coupling constant reaches values up to $3$\,kHz, an order-of-magnitude enhancement over the state-of-the-art. We estimate a spin detection sensitivity of $12 \pm 3$~spins$/\sqrt{\text{Hz}}$ for donors in silicon at millikelvin temperatures. As a counterpart, the Rabi frequency is highly inhomogeneous. Also, due to differential thermal contractions between the substrate and the nanowire, the donor resonance is considerably broadened by strain. 

\begin{figure*}
	\center
	\includegraphics[width=2\columnwidth]{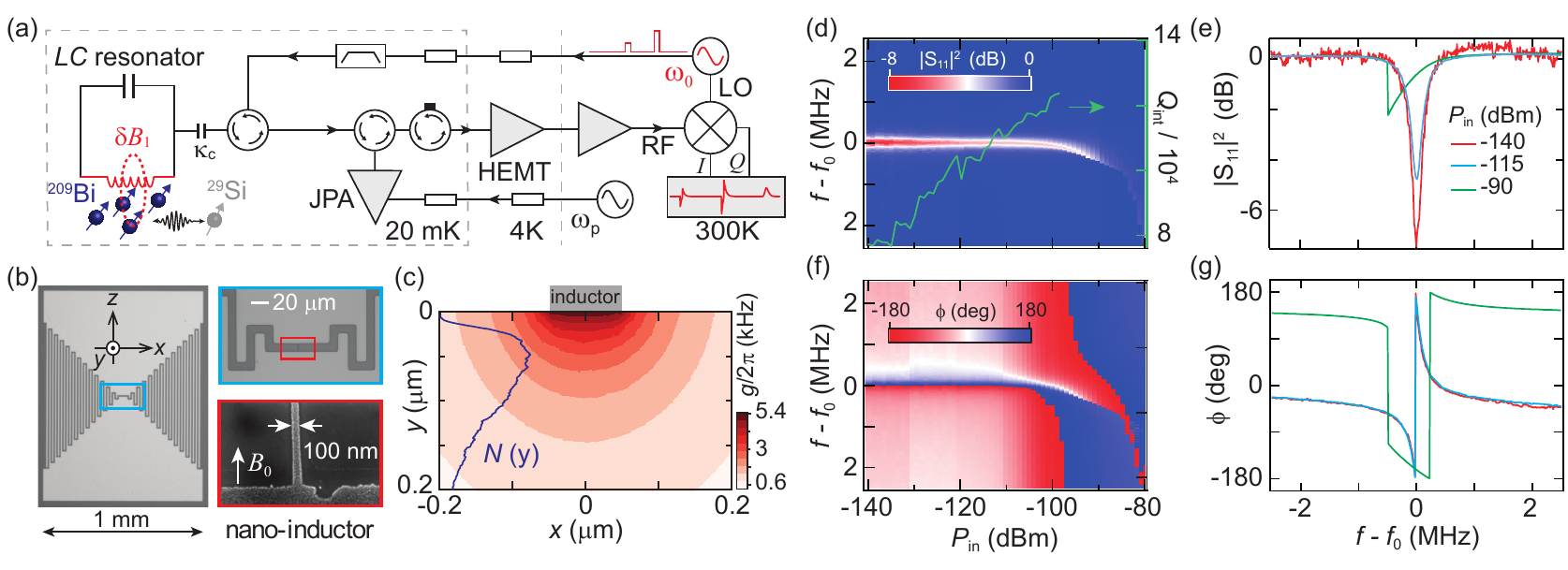}
	\caption{Quantum limited spectrometer. (a) Schematics of the measurement setup. Square control pulses are applied at resonator frequency $\omega_0$. Reflected and emitted signals from spins are amplified by a JPA in its degenerate mode, i.e. when pumped at $\omega_p \approx 2\omega_0$. (b) Optical and scanning electron microscope images of the $LC$ resonator with a nanometric inductor. Light and dark areas form the aluminium and the underlying Si substrate, respectively. The magnetic field is applied parallel to the inductor. (c) Calculated spin-resonator coupling strength distribution for the first spin-transition. Implantation profile $N (y)$ with peak concentration of $8\times 10^{16}~\text{cm}^{-3}$ is also plotted. Reflected power (d,e) and phase (f,g) from continuous wave measurements for resonator S1 at different input powers $P_\text{in}$. Extracted internal quality factor $Q_\text{int}$ is plotted on the right axis of panel (d). }
	\label{fig:setup}
\end{figure*}

An overview of our resonator design is shown in Fig.~\ref{fig:setup}(b). The resonator is patterned out of a 50~nm-thick superconducting aluminum film. It consists of an interdigitated capacitor $C$ with fingers $10~\mu$m wide and separated by the same amount. The bowtie shape is chosen to minimize the stray inductance of this capacitor~\cite{eichler_electron_2017}. The resulting resonator impedance is estimated from electromagnetic simulations~\footnote{CST Microwave Studios$\textsuperscript{\tiny\textregistered}$} to be $Z_c \sim 15~\Omega$. The capacitor is shorted by a $100~\text{nm}$-wide, $10~ \mu \mathrm{m}$-long wire that constitutes the resonator inductance $L$, and around which the magnetic component of the microwave field is by far the strongest. 
%This is the spin detection volume.  and the magnetic mode volume to be $\approx 6$~fL~\cite{haroche_exploring_2006}.

%%%%%%%%%%%%%%%%%%%%%%%%%%%%%%%%%%%%%%%%%%%%%%%%%%%%
\begin{figure}[t!]
	\center
	\includegraphics[width=\columnwidth]{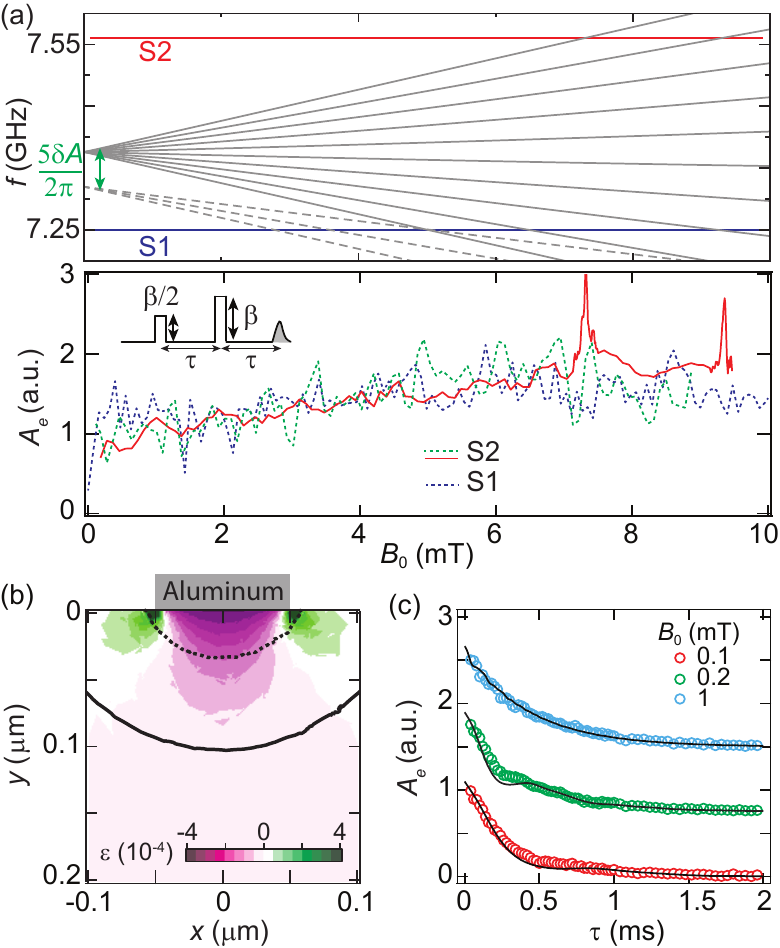}
	\caption{Spin spectroscopy (a) Top: Ten electron spin transitions of bismuth in the low magnetic field regime. The dashed lines represent new frequencies for first three transitions due to changes in hyperfine constant $\delta A$ by local strain. Bottom: Magnetic field sweep of the echo signal for two devices fabricated on the same substrate (sequence in inset). Different curves are scaled in amplitude and acquired at different pulse amplitudes $\beta$. For device S2, $\beta = 2\times 10^5~ \text{s}^{-1/2}$ (dashed curve), $7\times 10^5~\text{s}^{-1/2}$ and the pulse duration $dt = 0.5~\mu$s. For device S1, $\beta = 10^5~ \text{s}^{-1/2}$ and $dt = 1~\mu$s. (b) Hydrostatic component $\epsilon$ of the strain computed using finite element model simulations of COMSOL software. Solid and dashed curves represent respective contours of probed spins in S2 when assuming perfect $\pi$ pulses for two $\beta$ values.  (c) Two pulse spin coherence measurements with S1: Echo area $A_\text{s}$ as a function of delay $\tau$. Curves have been offset for clarity. Solid curves are simulations.}
	\label{fig:spectroscopy}
\end{figure}

%%%%%%%%%%%%%%%%%%%%%%%%%%%%%%%%%%%%%%%%%%%%%%%%%%%%%
The resonator is patterned on top of a silicon substrate in which bismuth atoms were implanted between $50$ and $100$\,nm below the surface (see Fig.~\ref{fig:setup}). To increase the donor coherence time, the substrate was enriched in the nuclear-spin-free $^{28}\text{Si}$ isotope, with a nominal residual $^{29}\text{Si}$ relative concentration of $5\times 10^{-4}$. Bismuth is an electron donor in silicon, and can trap the unpaired electron at low temperatures, whose electron spin $S=1/2$ provides the ESR signal~\cite{feher_electron_1959}. The donor spin Hamiltonian is $H/\hbar =\gamma_e \bold{S} \cdot \bold{B_0} + A \bold{S} \cdot \bold{I}$, where $I=9/2$ is the nuclear spin of the bismuth ion, $A/2\pi = 1.45~\text{GHz}$ the hyperfine coupling, and $B_0$ the value of a dc magnetic field applied parallel to the nanowire inductor (along the z axis). At low magnetic fields $B_0$, the ESR-allowed transitions~\cite{morley_initialization_2010,mohammady_bismuth_2010,george_electron_2010,wolfowicz_atomic_2013} are shown in Fig.~\ref{fig:spectroscopy}(a). Note the zero-field splitting of 7.38~GHz enabling measurements at low fields ($<10$\,mT in this work). 

A schematic description of the setup is shown in Fig.~\ref{fig:setup}(a). The sample is cooled at 20~mK, and probed by microwave reflectometry. Sequences of microwave pulses at the resonator frequency are applied to the sample input to drive the spins. The reflected pulses and emitted echo-signals are routed via a cryogenic circulator to the detection chain, consisting of a Josephson Parametric Amplifier (JPA)~\cite{zhou_high-gain_2014} at 20~mK followed by a High-Electron-Mobility Transistor (HEMT) at the 4K stage. The signal quadratures are then obtained by homodyne demodulation at room temperature. A phase cycling scheme is applied: we subtract two consecutive traces with the first pulse phase being changed from $+\pi/2$ to $-\pi/2$, which minimizes the impact of setup drifts. More details can be found in Ref.~\onlinecite{bienfait_reaching_2015,probst_inductive-detection_2017}. Results from two nearly identical samples (S1 and S2) are reported below; unless mentioned explicitly, measurements reported have been performed on sample S1.

%The resonator parameters are first determined using Vector Network Analyzer measurements, shown as a function of the input power for S1 in Fig.~\ref{fig:setup}(d). 
The resonator parameters are first determined by measuring its complex reflection coefficient $S_\text{11}$ with a vector network anlyzer (VNA); it is shown for S1 in Fig.~\ref{fig:setup}(d,f), as a function of the input power. At the lowest power ($P_\text{in}=-140$\,dBm), the resonance is well fitted by a Lorentzian, yielding a frequency $\omega_0 /2\pi = 7.25$\,GHz, an external quality factor $Q_\mathrm{ext} = 3 \times 10^4$, and an internal quality factor $Q_\mathrm{int} = 8 \times 10^4$. At larger power, the internal quality factor is observed to progressively increase, up to $Q_\mathrm{int}=1.2 \times 10^5$ for $P_\text{in}=-100$\,dBm. Such power-dependent internal losses are characteristic of dielectric losses caused by two-level systems and are commonly observed in superconducting micro-resonators \cite{martinis_decoherence_2005}. From $\sim -100$\,dBm to $\sim -90$\,dBm, the resonance appears to shift to lower frequencies, and its response in amplitude deviates progressively from a Lorentzian; at powers above $-90$\,dBm it shows abrupt changes as seen in Fig.~\ref{fig:setup}(e,g). Similar results are obtained for sample S2, with a frequency of $7.56$\,GHz and a total quality factor of $6\times 10^3$ dominated by external coupling. Such behavior is characteristic of the Kerr non-linearity caused by the nanowire kinetic inductance (which can be harnessed to achieve parametric amplification of microwave signals \cite{levenson-falk_nonlinear_2011}). For pulsed ESR spectroscopy, microwave pulses at $\omega_0$ are applied to drive the spins. Given the resonator non-linearity, it is preferable to use drive pulses with maximum input power lower than $-100$~dBm. With a slightly larger input power (between $\sim -100$~dBm and $\sim -90$~dBm), the intra-cavity field acquires a deterministic time-dependent phase shift, which causes the resulting echo signal to be non-trivially distributed on both the in-phase and out-of-phase quadratures; the echo magnitude itself is however little affected and this weakly non-linear regime can thus also be used for pulse ESR spectroscopy~\cite{asfaw_multi-frequency_2017}.

An important parameter is the spin-photon coupling strength $g_0 = \gamma_e \langle 0 | S_x | 1 \rangle \delta B_1$, defined as half the Rabi frequency that a spin would undergo in $\delta B_1$, the microwave amplitude corresponding to a 1 photon field, $\gamma_e/2\pi = 28$\,GHz/T being the free electron gyromagnetic ratio and $\langle 0 | S_x | 1 \rangle$ the spin transition matrix element between the two levels $|0\rangle$ and $|1\rangle$ whose frequency difference is equal to $\omega_0$ at the applied $B_0$. The spin-photon coupling strength for the lowest-frequency bismuth donor transition is shown in Fig.~\ref{fig:setup}(c). The magnetic field $\delta B_1$ was calculated using an electromagnetic solver assuming that the ac current corresponding to 1 photon in the resonator $\delta i = \omega_0 \sqrt{\hbar / 2 Z_c}$ flows through the inductor. Close to the nanowire within a depth of 40~nm, $g_0/2\pi$ reaches values as high as $3-5$~kHz, one order of magnitude larger than spin-resonator couplings measured so far\cite{probst_inductive-detection_2017}, thanks to the extreme confinement of the microwave field around the nanowire. The price to pay however is a large spatial inhomogeneity of $g_0$. Consequences of a wide distribution of $g_0$ values when combined with spin relaxation by the Purcell effect were analyzed in Ref.~\onlinecite{ranjan_pulsed_2020} for typical pulse ESR sequences. Consider a two-pulse-echo sequence, consisting of a first microwave pulse of amplitude $\beta/2$ and duration $dt$, followed by a waiting time $\tau$, and by a second pulse of amplitude $\beta$ and duration $dt$ [see Fig.~\ref{fig:spectroscopy}(a)] with $\beta = \sqrt{P_\text{in}/\hbar \omega_0}$ ($\beta^2$ represents number of incoming photons per second). Because of the $g_0$ inhomogeneous distribution, the spin echo signal observed a time $\tau$ after the second pulse receives its dominant contribution from spins with a coupling constant $g_0(\beta) = \pi \sqrt{\kappa} / (4\beta dt)$, because those undergo Rabi angles close to $\pi/2$ and $\pi$. It is therefore possible to probe spins with different coupling constants by changing the amplitude $\beta$ of the detection echo sequence~\cite{ranjan_pulsed_2020}. 

A first example of this selectivity is provided by spectrum measurements shown in Fig.~\ref{fig:spectroscopy}(a). The integral $A_\text{e}$ of a spin-echo is displayed as a function of the value of the magnetic field $B_0$, applied parallel to the inductor. Two spectra measured in S2 are displayed, for two values of $\beta$ (corresponding to input powers of -100 dBm and -86 dBm). In the high-power spectrum, two narrow peaks (solid line) are observed close to the expected bismuth donor ESR transitions, on top of an approximately constant signal that extends until $B_0=0$. In the low-power spectrum, the peaks vanish and only the nearly constant echo signal is observed. Only the low-power curve was measured with resonator S1, and a spectrum similar to the one of sample S2 is observed [see Fig.~\ref{fig:spectroscopy}(a)].

These observations suggest that bismuth donors spins closest to the wire (those detected in the low-power measurement) have a very broad spectrum, whereas those far from the wire (detected in the high-power measurement) have narrower linewidth. This can be qualitatively understood by the effect of mechanical strain on the spin properties of bismuth donors. The hyperfine constant was recently shown~\cite{mansir_linear_2018} to depend linearly on the hydrostatic strain $\epsilon$, with $(dA/d\epsilon)/2 \pi \sim 29$~GHz. Aluminum contracts $10$ times more than silicon upon cooldown from room-temperature to $10$\,mK. The calculated strain profile $\epsilon$ in the silicon resulting from the differential thermal contraction of the aluminum inductor patterned on top is shown in Fig.~\ref{fig:spectroscopy}(b)~\cite{pla_strain-induced_2018}. For spins located in the region close to the inductance (and therefore strongly coupled to the resonator), the standard-deviation in the zero-field splitting (equal to $5A/2\pi$) is $\sim 100$\,MHz, which is sufficient to account for a complete overlap of neighboring peaks and therefore a nearly flat spectrum. Spins further from the inductor are submitted to much lower strain, leading to better-resolved transitions.

One consequence of this broadening for donors near the inductor is that each of the $10$ Bismuth transitions may contribute to the spin-echo signal measured at a given $B_0$, as schematically explained in Fig.~\ref{fig:spectroscopy}(a). We take that into account in our analysis as explained below.

To examine that the echo signals arise from implanted bismuth donors and not from surface impurities, we get information about the electron spin environment using hyperfine spectroscopy. The integrated spin-echo amplitude $A_e$ is measured as a function of $\tau$ for various values of $B_0$ [see Fig.~\ref{fig:spectroscopy}(c)]. For $B_0 \geq 1$\,mT, the decay is exponential, with a decay constant $T_2= 0.85\pm 0.1$~ms, a coherence time somewhat shorter than comparable measurements in identical silicon samples in refs.~\cite{bienfait_reaching_2015,probst_inductive-detection_2017}. At $B_0<1$\,mT, slight oscillations are visible. They approximately match the expected Electron Spin Echo Envelope Modulation (ESEEM) by a bath of $^{29}$Si nuclei with the nominal relative concentration of $5 \times 10^{-4}$ [see Fig.~\ref{fig:spectroscopy}(c)] ~\cite{probst_hyperfine_2020}. This suggests that the spin-echo signals result from spins located in the bulk of the silicon sample.

\begin{figure}[t!]
	\center
	\includegraphics[width=\columnwidth]{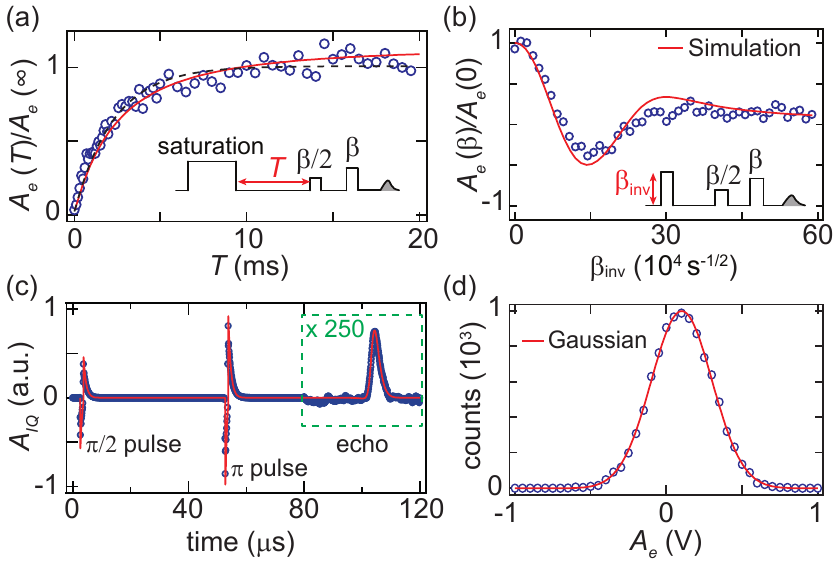}
	\caption{Spectrometer sensitivity. (a) Spin relaxation time measured using the saturation recovery method. The dashed curve is the best fit to an exponential yielding $T_1=2\pm 0.4$~ms.  (b) Rabi oscillations performed using three pulse sequence shown in the inset at $\grep=50$~Hz.  (c) Measurement of the reflected amplitude $A_{IQ}$ showing a complete Hahn echo sequence. (d) Histograms of echoes at $\grep = 100~$Hz measured with phase cycling. Solid curves in panels (a-c) are numerical simulations taking into account all ten ESR transitions. }
	\label{fig:spectrometer}
\end{figure}

In the Purcell regime, $T_1 = \kappa / (4g_0^2)$ at resonance, implying that the spin-photon coupling constant can be deduced from spin relaxation measurements. The corresponding spin relaxation sequence (saturation recovery) consists of a long saturation pulse, followed after a delay $T$ by a detection-echo whose pulse amplitude $\beta_0 = 6\times 10^4~\text{s}^{-1/2}$ ($P_\text{in} \approx -107 $~dBm, thus in the resonator linear regime) mostly selects a class of spins with coupling constant $g_0(\beta_0)$. The integrated echo is shown in Fig.~\ref{fig:spectrometer}(a) as a function of $T$. An exponential fit yields $T_1 = 2 \pm 0.4$\,ms, which translates into a coupling constant $g_0(\beta_0)/2\pi = 2.7$\,kHz. This is the largest spin-photon magnetic coupling measured, confirming the predicted coupling distribution [Fig.~\ref{fig:setup}(c)]. Larger values may be obtained using superconducting flux-qubits\cite{zhu_coherent_2011}. As discussed in Ref.~\onlinecite{ranjan_pulsed_2020}, the measured relaxation time scales as $\beta^2$, in good agreement with simulations. This enables us to calibrate $\beta$ in absolute units, which is otherwise difficult because of the imperfect knowledge of the total attenuation in the input line. The validity of the calibration is confirmed by a Rabi nutation experiment, measured with a pulse sequence shown in Fig.~\ref{fig:spectrometer}(b). The integrated echo area is shown as a function of the amplitude $\beta_\text{inv}$ of a first microwave pulse, followed by a detection echo. The simulation agrees quantitatively with the data, without adjustable parameter. 

We then estimate the spectrometer sensitivity following the method explained in Ref.~\onlinecite{probst_inductive-detection_2017}. The number of spins $N_\text{spin}$ contributing to an echo signal, defined as the total number of spins excited after the initial pulse is first determined: we measure a complete Hahn-echo sequence, including two control pulses of amplitude $\beta_0/2, \beta_0$ and duration $1~\mu$s separated by $50~\mu \text{s}$. The JPA was switched off to avoid its saturation during application of the control pulses. The reflected amplitude, obtained after $10^6$ averages measured with a repetition rate of $100$~Hz, is shown in Fig.~\ref{fig:spectrometer}(c). Since the ratio between echo and control pulse amplitude is uniquely determined by $N_\text{spin}$, the latter is obtained by adjusting the simulations to best fit the data [see solid line in Fig.~\ref{fig:spectrometer}(c)], yielding $N_\text{spin} \approx 36 \pm 8$. We note that all ten transitions are equally weighted to account for the overlap of the ESR transition due to strain-induced large spectral broadening. This number is also roughly consistent with the number of bismuth atoms expected in the resonator magnetic mode volume, taking into account that the spin resonance linewidth is considerably broader than the detection bandwidth.

The signal-to-noise ratio (SNR) is measured by acquiring $10^4$ echo traces (with phase cycling) in the degenerate mode of the JPA at $\grep = 100~$Hz. From the histogram shown in Fig.~\ref{fig:spectrometer}(d), we find that the SNR is 0.33 for a single echo trace. Therefore, one could detect $\Nmin \approx 120 \pm 24$ spins with unity SNR in a single echo sequence. Since the measurements are repeated with $100$\,Hz repetition rate, this translates into a spin detection sensitivity of $12\pm 3$~spins$/\sqrt{\text{Hz}}$. Theoretical estimates of the sensitivity~\cite{bienfait_reaching_2015} predict that $\Nmin=\frac{\kappa} {2 P g_0} \sqrt{\tilde{n}}$, $\tilde{n}=1/2$ being the noise due to quantum fluctuations of the microwave field and $P \approx 1$, the thermal polarization of the spins at 20~mK. For $g_0/2\pi=2.7~$kHz, this yields $\Nmin = 50$, in semi-quantitative agreement with the measured value. Further improvements in sensitivity would require to further increase $g_0$ or the quality factor, or to reduce the noise below the quantum limit using squeezed vacuum for instance, as demonstrated recently~\cite{bienfait_magnetic_2017}.

\begin{figure}[t!]
	\center
	\includegraphics[width=\columnwidth]{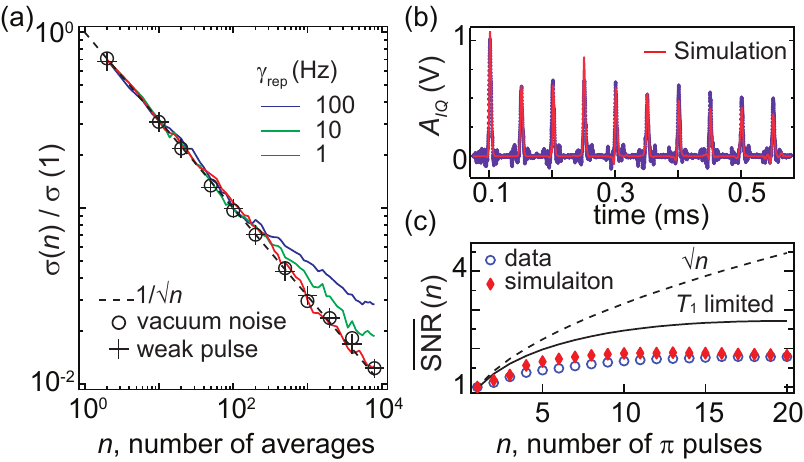}
	\caption{Correlated noise and CPMG sequences (a) Measured standard deviation of signals versus number $n$ of averages shown by solid lines and symbols. (b) In symbols, averaged amplitude of $10^4$ repetitions of CPMG traces measured at $\grep=100~$Hz in the degenerate mode of the JPA. Refocusing pulses are not visible due to phase cycling. (c) Improvement of echo SNR with number of refocusing $\pi$ pulses. Simulated SNR (diamond symbols), and calculated curves for cases of no energy relaxation (dashed line) and finite relaxation (solid line) assuming uncorrelated noise between echoes. }
	\label{fig:CPMG}
\end{figure}

Expressing the spin sensitivity in spin/$\sqrt{\mathrm{Hz}}$ assumes that repeating the same sequence $n$ times and averaging the result increases the SNR by $\sqrt{n}$. We test this assumption by acquiring $10^7$ echo sequences, repeated with a rate of $100$\,Hz, generating a histogram obtained by averaging $n$ consecutive echo integrals, and computing the standard deviation $\sigma(n)$. The result is shown in Fig.~\ref{fig:CPMG}(a). We observe that, while $\sigma(n)$ indeed scales like $1/\sqrt{n}$ until $n=200$, it keeps going down for larger values of $n$ but slower than $1/\sqrt{n}$.

To test whether the deviation of $\sigma(n)$ from $1/\sqrt{n}$ is due to the setup or to the sample, we mimic the echo acquisition by sending a train of weak coherent pulses with an amplitude that corresponds to an echo and the same repetition rate of 100~Hz at the resonator frequency $\omega_0$. Their standard deviation now follows the $1/\sqrt{n}$ law until at least $n=10^4$, implying that the slower-than-$\sqrt{n}$ echo averaging is not due to setup drift. Note that compared (and contrary) to the analysis performed in Ref.~\onlinecite{probst_inductive-detection_2017}, the test pulses were sent at the resonator frequency $\omega_0$ so that they would be affected by resonator phase noise, which can thus be ruled out as the origin of the slower-than-$\sqrt{n}$ echo signal averaging. 

Further insight is obtained by analysing the spin-echo data differently: instead of averaging $n$ consecutive echo traces, we average them with a separation of $10$, or $100$ traces (which amounts to effectively changing the repetition rate to $10$~Hz or $1$\,Hz). As seen in Fig.~\ref{fig:CPMG}(a), the $1/\sqrt{n}$ law is progressively recovered. A possible interpretation is that the number of spins contributing to the echo slightly fluctuates over a time scale of a few seconds, possibly due to a slow redistribution of the bismuth donor population within the hyperfine states or to ionization/neutralization dynamics of one or a few donors located close to the metallic electrodes.

One way to increase the spectrometer SNR is to add extra refocusing pulses after the emission of the Hahn echo in order to obtain several echoes per sequence~\cite{mentink-vigier_increasing_2013,bienfait_reaching_2015}. To this end, we use a Carr-Purcell-Meiboom-Gill (CPMG) sequence: $(\pi/2)_{\pm x}-\tau-\pi_y-\tau-\left(-echo-\tau/2-\pi_y-\tau/2-\right)_{n}-echo$. The echo train generated by this sequence is shown as symbols in Fig.~\ref{fig:CPMG}(b) for $\tau=50~\mu$s (the refocusing pulses are not visible due to phase cycling). The numerical simulation plotted as a solid line describes well the change in the amplitude over time, without any adjustable parameter other than the overall amplitude. Note that the echo amplitude quickly decays after the first pulse, due to the large pulse errors caused by the $B_1$ inhomogeneity. In order to quantify the SNR improvement, we measure $10^4$ sequences of CPMG echoes (with a repetition rate $\grep = 100~$Hz). We then generate histograms obtained by averaging the first $n$ echoes of each sequence, divide the mean by the standard deviation, yielding the SNR as a function of $n$. We find a maximum SNR improvement of $\approx 2$, which corresponds to a spin detection sensitivity of 6~spins$/\sqrt{\text{Hz}}$. This enhancement is well reproduced by simulations and is not far from the maximum limit $\sim 2.7$ set by the energy relaxation [see Fig.~\ref{fig:CPMG}(c)]. 

%conclusion
In summary we demonstrate a sensitivity of 12~spins$/\sqrt{\text{Hz}}$ in inductively-detected ESR spectroscopy, using a resonator based on a superconducting nanowire with a $6$~fL detection volume. The spin-resonator coupling reaches values up to $3$\,kHz. An ESEEM signal originating from $\sim 30$ electron spins coupled to residual $^{29}$Si nuclear spins was detected. Future work will study possible applications of superconducting ESR spectroscopy to real-world systems, for instance paramagnetic defects in two-dimensional Van der Waals materials~\cite{geim_van_2013}.

We acknowledge technical support from P.~S\'enat and P.-F.~Orfila, as well as stimulating discussions within the Quantronics group. We acknowledge support of the European Research Council under the European Community's Seventh Framework Programme (FP7/2007-2013) through grant agreements No.~615767 (CIRQUSS), 279781 (ASCENT), and 630070 (quRAM), of the ANR projects QIPSE and NASNIQ (contract number ANR-17-CHIN-0001). T. S. was supported by the U.S. Department of Energy under Contract No. DE-AC02-05CH11231.

%% Bibliography
%\bibliography{FemtoliterEPR}

\begin{thebibliography}{38}%
\makeatletter
\providecommand \@ifxundefined [1]{%
 \@ifx{#1\undefined}
}%
\providecommand \@ifnum [1]{%
 \ifnum #1\expandafter \@firstoftwo
 \else \expandafter \@secondoftwo
 \fi
}%
\providecommand \@ifx [1]{%
 \ifx #1\expandafter \@firstoftwo
 \else \expandafter \@secondoftwo
 \fi
}%
\providecommand \natexlab [1]{#1}%
\providecommand \enquote  [1]{``#1''}%
\providecommand \bibnamefont  [1]{#1}%
\providecommand \bibfnamefont [1]{#1}%
\providecommand \citenamefont [1]{#1}%
\providecommand \href@noop [0]{\@secondoftwo}%
\providecommand \href [0]{\begingroup \@sanitize@url \@href}%
\providecommand \@href[1]{\@@startlink{#1}\@@href}%
\providecommand \@@href[1]{\endgroup#1\@@endlink}%
\providecommand \@sanitize@url [0]{\catcode `\\12\catcode `\$12\catcode
  `\&12\catcode `\#12\catcode `\^12\catcode `\_12\catcode `\%12\relax}%
\providecommand \@@startlink[1]{}%
\providecommand \@@endlink[0]{}%
\providecommand \url  [0]{\begingroup\@sanitize@url \@url }%
\providecommand \@url [1]{\endgroup\@href {#1}{\urlprefix }}%
\providecommand \urlprefix  [0]{URL }%
\providecommand \Eprint [0]{\href }%
\providecommand \doibase [0]{http://dx.doi.org/}%
\providecommand \selectlanguage [0]{\@gobble}%
\providecommand \bibinfo  [0]{\@secondoftwo}%
\providecommand \bibfield  [0]{\@secondoftwo}%
\providecommand \translation [1]{[#1]}%
\providecommand \BibitemOpen [0]{}%
\providecommand \bibitemStop [0]{}%
\providecommand \bibitemNoStop [0]{.\EOS\space}%
\providecommand \EOS [0]{\spacefactor3000\relax}%
\providecommand \BibitemShut  [1]{\csname bibitem#1\endcsname}%
\let\auto@bib@innerbib\@empty
%</preamble>
\bibitem [{\citenamefont {Schweiger}\ and\ \citenamefont
  {Jeschke}(2001)}]{schweiger_principles_2001}%
  \BibitemOpen
  \bibfield  {author} {\bibinfo {author} {\bibfnamefont {A.}~\bibnamefont
  {Schweiger}}\ and\ \bibinfo {author} {\bibfnamefont {G.}~\bibnamefont
  {Jeschke}},\ }\href@noop {} {\emph {\bibinfo {title} {Principles of pulse
  electron paramagnetic resonance}}}\ (\bibinfo  {publisher} {Oxford University
  Press},\ \bibinfo {year} {2001})\BibitemShut {NoStop}%
\bibitem [{\citenamefont {Wrachtrup}\ \emph {et~al.}(1993)\citenamefont
  {Wrachtrup}, \citenamefont {Von~Borczyskowski}, \citenamefont {Bernard},
  \citenamefont {Orritt},\ and\ \citenamefont
  {Brown}}]{wrachtrup_optical_1993}%
  \BibitemOpen
  \bibfield  {author} {\bibinfo {author} {\bibfnamefont {J.}~\bibnamefont
  {Wrachtrup}}, \bibinfo {author} {\bibfnamefont {C.}~\bibnamefont
  {Von~Borczyskowski}}, \bibinfo {author} {\bibfnamefont {J.}~\bibnamefont
  {Bernard}}, \bibinfo {author} {\bibfnamefont {M.}~\bibnamefont {Orritt}}, \
  and\ \bibinfo {author} {\bibfnamefont {R.}~\bibnamefont {Brown}},\ }\href
  {\doibase 10.1038/363244a0} {\bibfield  {journal} {\bibinfo  {journal}
  {Nature}\ }\textbf {\bibinfo {volume} {363}},\ \bibinfo {pages} {244}
  (\bibinfo {year} {1993})}\BibitemShut {NoStop}%
\bibitem [{\citenamefont {Gruber}\ \emph {et~al.}(1997)\citenamefont {Gruber},
  \citenamefont {Dräbenstedt}, \citenamefont {Tietz}, \citenamefont {Fleury},
  \citenamefont {Wrachtrup},\ and\ \citenamefont
  {Borczyskowski}}]{gruber_scanning_1997}%
  \BibitemOpen
  \bibfield  {author} {\bibinfo {author} {\bibfnamefont {A.}~\bibnamefont
  {Gruber}}, \bibinfo {author} {\bibfnamefont {A.}~\bibnamefont
  {Dräbenstedt}}, \bibinfo {author} {\bibfnamefont {C.}~\bibnamefont {Tietz}},
  \bibinfo {author} {\bibfnamefont {L.}~\bibnamefont {Fleury}}, \bibinfo
  {author} {\bibfnamefont {J.}~\bibnamefont {Wrachtrup}}, \ and\ \bibinfo
  {author} {\bibfnamefont {C.~v.}\ \bibnamefont {Borczyskowski}},\ }\href
  {\doibase 10.1126/science.276.5321.2012} {\bibfield  {journal} {\bibinfo
  {journal} {Science}\ }\textbf {\bibinfo {volume} {276}},\ \bibinfo {pages}
  {2012} (\bibinfo {year} {1997})}\BibitemShut {NoStop}%
\bibitem [{\citenamefont {Rugar}, \citenamefont {Yannoni},\ and\ \citenamefont
  {Sidles}(1992)}]{rugar_mechanical_1992}%
  \BibitemOpen
  \bibfield  {author} {\bibinfo {author} {\bibfnamefont {D.}~\bibnamefont
  {Rugar}}, \bibinfo {author} {\bibfnamefont {C.}~\bibnamefont {Yannoni}}, \
  and\ \bibinfo {author} {\bibfnamefont {J.}~\bibnamefont {Sidles}},\ }\href
  {\doibase 10.1038/360563a0} {\bibfield  {journal} {\bibinfo  {journal}
  {Nature}\ }\textbf {\bibinfo {volume} {360}},\ \bibinfo {pages} {563}
  (\bibinfo {year} {1992})}\BibitemShut {NoStop}%
\bibitem [{\citenamefont {Rugar}\ \emph {et~al.}(2004)\citenamefont {Rugar},
  \citenamefont {Budakian}, \citenamefont {Mamin},\ and\ \citenamefont
  {Chui}}]{rugar_single_2004}%
  \BibitemOpen
  \bibfield  {author} {\bibinfo {author} {\bibfnamefont {D.}~\bibnamefont
  {Rugar}}, \bibinfo {author} {\bibfnamefont {R.}~\bibnamefont {Budakian}},
  \bibinfo {author} {\bibfnamefont {H.}~\bibnamefont {Mamin}}, \ and\ \bibinfo
  {author} {\bibfnamefont {B.}~\bibnamefont {Chui}},\ }\href {\doibase
  10.1038/nature02658} {\bibfield  {journal} {\bibinfo  {journal} {Nature}\
  }\textbf {\bibinfo {volume} {430}},\ \bibinfo {pages} {329} (\bibinfo {year}
  {2004})}\BibitemShut {NoStop}%
\bibitem [{\citenamefont {Chamberlin}, \citenamefont {Moberly},\ and\
  \citenamefont {Symko}(1979)}]{chamberlin_high-sensitivity_1979}%
  \BibitemOpen
  \bibfield  {author} {\bibinfo {author} {\bibfnamefont {R.~V.}\ \bibnamefont
  {Chamberlin}}, \bibinfo {author} {\bibfnamefont {L.~A.}\ \bibnamefont
  {Moberly}}, \ and\ \bibinfo {author} {\bibfnamefont {O.~G.}\ \bibnamefont
  {Symko}},\ }\href {\doibase 10.1007/BF00115584} {\bibfield  {journal}
  {\bibinfo  {journal} {Journal of Low Temperature Physics}\ }\textbf {\bibinfo
  {volume} {35}},\ \bibinfo {pages} {337} (\bibinfo {year} {1979})}\BibitemShut
  {NoStop}%
\bibitem [{\citenamefont {Manassen}\ \emph {et~al.}(1989)\citenamefont
  {Manassen}, \citenamefont {Hamers}, \citenamefont {Demuth},\ and\
  \citenamefont {Castellano~Jr.}}]{manassen_direct_1989}%
  \BibitemOpen
  \bibfield  {author} {\bibinfo {author} {\bibfnamefont {Y.}~\bibnamefont
  {Manassen}}, \bibinfo {author} {\bibfnamefont {R.~J.}\ \bibnamefont
  {Hamers}}, \bibinfo {author} {\bibfnamefont {J.~E.}\ \bibnamefont {Demuth}},
  \ and\ \bibinfo {author} {\bibfnamefont {A.~J.}\ \bibnamefont
  {Castellano~Jr.}},\ }\href {\doibase 10.1103/PhysRevLett.62.2531} {\bibfield
  {journal} {\bibinfo  {journal} {Phys. Rev. Lett.}\ }\textbf {\bibinfo
  {volume} {62}},\ \bibinfo {pages} {2531} (\bibinfo {year}
  {1989})}\BibitemShut {NoStop}%
\bibitem [{\citenamefont {Baumann}\ \emph {et~al.}(2015)\citenamefont
  {Baumann}, \citenamefont {Paul}, \citenamefont {Choi}, \citenamefont {Lutz},
  \citenamefont {Ardavan},\ and\ \citenamefont
  {Heinrich}}]{baumann_electron_2015}%
  \BibitemOpen
  \bibfield  {author} {\bibinfo {author} {\bibfnamefont {S.}~\bibnamefont
  {Baumann}}, \bibinfo {author} {\bibfnamefont {W.}~\bibnamefont {Paul}},
  \bibinfo {author} {\bibfnamefont {T.}~\bibnamefont {Choi}}, \bibinfo {author}
  {\bibfnamefont {C.~P.}\ \bibnamefont {Lutz}}, \bibinfo {author}
  {\bibfnamefont {A.}~\bibnamefont {Ardavan}}, \ and\ \bibinfo {author}
  {\bibfnamefont {A.~J.}\ \bibnamefont {Heinrich}},\ }\href {\doibase
  10.1126/science.aac8703} {\bibfield  {journal} {\bibinfo  {journal}
  {Science}\ }\textbf {\bibinfo {volume} {350}},\ \bibinfo {pages} {417}
  (\bibinfo {year} {2015})}\BibitemShut {NoStop}%
\bibitem [{\citenamefont {Morello}\ \emph {et~al.}(2010)\citenamefont
  {Morello}, \citenamefont {Pla}, \citenamefont {Zwanenburg}, \citenamefont
  {Chan}, \citenamefont {Tan}, \citenamefont {Huebl}, \citenamefont {Mottonen},
  \citenamefont {Nugroho}, \citenamefont {Yang}, \citenamefont {van
  Donkelaar},\ and\ \citenamefont {{others}}}]{morello_single-shot_2010}%
  \BibitemOpen
  \bibfield  {author} {\bibinfo {author} {\bibfnamefont {A.}~\bibnamefont
  {Morello}}, \bibinfo {author} {\bibfnamefont {J.~J.}\ \bibnamefont {Pla}},
  \bibinfo {author} {\bibfnamefont {F.~A.}\ \bibnamefont {Zwanenburg}},
  \bibinfo {author} {\bibfnamefont {K.~W.}\ \bibnamefont {Chan}}, \bibinfo
  {author} {\bibfnamefont {K.~Y.}\ \bibnamefont {Tan}}, \bibinfo {author}
  {\bibfnamefont {H.}~\bibnamefont {Huebl}}, \bibinfo {author} {\bibfnamefont
  {M.}~\bibnamefont {Mottonen}}, \bibinfo {author} {\bibfnamefont {C.~D.}\
  \bibnamefont {Nugroho}}, \bibinfo {author} {\bibfnamefont {C.}~\bibnamefont
  {Yang}}, \bibinfo {author} {\bibfnamefont {J.~A.}\ \bibnamefont {van
  Donkelaar}}, \ and\ \bibinfo {author} {\bibnamefont {{others}}},\ }\href
  {\doibase doi:10.1038/nature09392} {\bibfield  {journal} {\bibinfo  {journal}
  {Nature}\ }\textbf {\bibinfo {volume} {467}},\ \bibinfo {pages} {687}
  (\bibinfo {year} {2010})}\BibitemShut {NoStop}%
\bibitem [{\citenamefont {Narkowicz}, \citenamefont {Suter},\ and\
  \citenamefont {Niemeyer}(2008)}]{narkowicz_scaling_2008}%
  \BibitemOpen
  \bibfield  {author} {\bibinfo {author} {\bibfnamefont {R.}~\bibnamefont
  {Narkowicz}}, \bibinfo {author} {\bibfnamefont {D.}~\bibnamefont {Suter}}, \
  and\ \bibinfo {author} {\bibfnamefont {I.}~\bibnamefont {Niemeyer}},\ }\href
  {\doibase 10.1063/1.2964926} {\bibfield  {journal} {\bibinfo  {journal}
  {Review of Scientific Instruments}\ }\textbf {\bibinfo {volume} {79}},\
  \bibinfo {pages} {084702} (\bibinfo {year} {2008})}\BibitemShut {NoStop}%
\bibitem [{\citenamefont {Artzi}, \citenamefont {Twig},\ and\ \citenamefont
  {Blank}(2015)}]{artzi_induction-detection_2015}%
  \BibitemOpen
  \bibfield  {author} {\bibinfo {author} {\bibfnamefont {Y.}~\bibnamefont
  {Artzi}}, \bibinfo {author} {\bibfnamefont {Y.}~\bibnamefont {Twig}}, \ and\
  \bibinfo {author} {\bibfnamefont {A.}~\bibnamefont {Blank}},\ }\href
  {\doibase 10.1063/1.4913806} {\bibfield  {journal} {\bibinfo  {journal}
  {Applied Physics Letters}\ }\textbf {\bibinfo {volume} {106}},\ \bibinfo
  {pages} {084104} (\bibinfo {year} {2015})}\BibitemShut {NoStop}%
\bibitem [{\citenamefont {Sidabras}\ \emph {et~al.}(2019)\citenamefont
  {Sidabras}, \citenamefont {Duan}, \citenamefont {Winkler}, \citenamefont
  {Happe}, \citenamefont {Hussein}, \citenamefont {Zouni}, \citenamefont
  {Suter}, \citenamefont {Schnegg}, \citenamefont {Lubitz},\ and\ \citenamefont
  {Reijerse}}]{sidabras_extending_2019}%
  \BibitemOpen
  \bibfield  {author} {\bibinfo {author} {\bibfnamefont {J.~W.}\ \bibnamefont
  {Sidabras}}, \bibinfo {author} {\bibfnamefont {J.}~\bibnamefont {Duan}},
  \bibinfo {author} {\bibfnamefont {M.}~\bibnamefont {Winkler}}, \bibinfo
  {author} {\bibfnamefont {T.}~\bibnamefont {Happe}}, \bibinfo {author}
  {\bibfnamefont {R.}~\bibnamefont {Hussein}}, \bibinfo {author} {\bibfnamefont
  {A.}~\bibnamefont {Zouni}}, \bibinfo {author} {\bibfnamefont
  {D.}~\bibnamefont {Suter}}, \bibinfo {author} {\bibfnamefont
  {A.}~\bibnamefont {Schnegg}}, \bibinfo {author} {\bibfnamefont
  {W.}~\bibnamefont {Lubitz}}, \ and\ \bibinfo {author} {\bibfnamefont {E.~J.}\
  \bibnamefont {Reijerse}},\ }\href {\doibase 10.1126/sciadv.aay1394}
  {\bibfield  {journal} {\bibinfo  {journal} {Science Advances}\ }\textbf
  {\bibinfo {volume} {5}},\ \bibinfo {pages} {eaay1394} (\bibinfo {year}
  {2019})}\BibitemShut {NoStop}%
\bibitem [{\citenamefont {Malissa}\ \emph {et~al.}(2013)\citenamefont
  {Malissa}, \citenamefont {Schuster}, \citenamefont {Tyryshkin}, \citenamefont
  {Houck},\ and\ \citenamefont {Lyon}}]{malissa_superconducting_2013}%
  \BibitemOpen
  \bibfield  {author} {\bibinfo {author} {\bibfnamefont {H.}~\bibnamefont
  {Malissa}}, \bibinfo {author} {\bibfnamefont {D.~I.}\ \bibnamefont
  {Schuster}}, \bibinfo {author} {\bibfnamefont {A.~M.}\ \bibnamefont
  {Tyryshkin}}, \bibinfo {author} {\bibfnamefont {A.~A.}\ \bibnamefont
  {Houck}}, \ and\ \bibinfo {author} {\bibfnamefont {S.~A.}\ \bibnamefont
  {Lyon}},\ }\href {\doibase 10.1063/1.4792205} {\bibfield  {journal} {\bibinfo
   {journal} {Review of Scientific Instruments}\ }\textbf {\bibinfo {volume}
  {84}},\ \bibinfo {pages} {025116} (\bibinfo {year} {2013})}\BibitemShut
  {NoStop}%
\bibitem [{\citenamefont {Sigillito}\ \emph {et~al.}(2014)\citenamefont
  {Sigillito}, \citenamefont {Malissa}, \citenamefont {Tyryshkin},
  \citenamefont {Riemann}, \citenamefont {Abrosimov}, \citenamefont {Becker},
  \citenamefont {Pohl}, \citenamefont {Thewalt}, \citenamefont {Itoh},
  \citenamefont {Morton}, \citenamefont {Houck}, \citenamefont {Schuster},\
  and\ \citenamefont {Lyon}}]{sigillito_fast_2014}%
  \BibitemOpen
  \bibfield  {author} {\bibinfo {author} {\bibfnamefont {A.~J.}\ \bibnamefont
  {Sigillito}}, \bibinfo {author} {\bibfnamefont {H.}~\bibnamefont {Malissa}},
  \bibinfo {author} {\bibfnamefont {A.~M.}\ \bibnamefont {Tyryshkin}}, \bibinfo
  {author} {\bibfnamefont {H.}~\bibnamefont {Riemann}}, \bibinfo {author}
  {\bibfnamefont {N.~V.}\ \bibnamefont {Abrosimov}}, \bibinfo {author}
  {\bibfnamefont {P.}~\bibnamefont {Becker}}, \bibinfo {author} {\bibfnamefont
  {H.-J.}\ \bibnamefont {Pohl}}, \bibinfo {author} {\bibfnamefont {M.~L.~W.}\
  \bibnamefont {Thewalt}}, \bibinfo {author} {\bibfnamefont {K.~M.}\
  \bibnamefont {Itoh}}, \bibinfo {author} {\bibfnamefont {J.~J.~L.}\
  \bibnamefont {Morton}}, \bibinfo {author} {\bibfnamefont {A.~A.}\
  \bibnamefont {Houck}}, \bibinfo {author} {\bibfnamefont {D.~I.}\ \bibnamefont
  {Schuster}}, \ and\ \bibinfo {author} {\bibfnamefont {S.~A.}\ \bibnamefont
  {Lyon}},\ }\href {\doibase 10.1063/1.4881613} {\bibfield  {journal} {\bibinfo
   {journal} {Applied Physics Letters}\ }\textbf {\bibinfo {volume} {104}},\
  (\bibinfo {year} {2014})}\BibitemShut {NoStop}%
\bibitem [{\citenamefont {Bienfait}\ \emph {et~al.}(2015)\citenamefont
  {Bienfait}, \citenamefont {Pla}, \citenamefont {Kubo}, \citenamefont {Stern},
  \citenamefont {Zhou}, \citenamefont {Lo}, \citenamefont {Weis}, \citenamefont
  {Schenkel}, \citenamefont {Thewalt}, \citenamefont {Vion}, \citenamefont
  {Esteve}, \citenamefont {Julsgaard}, \citenamefont {Moelmer}, \citenamefont
  {Morton},\ and\ \citenamefont {Bertet}}]{bienfait_reaching_2015}%
  \BibitemOpen
  \bibfield  {author} {\bibinfo {author} {\bibfnamefont {A.}~\bibnamefont
  {Bienfait}}, \bibinfo {author} {\bibfnamefont {J.}~\bibnamefont {Pla}},
  \bibinfo {author} {\bibfnamefont {Y.}~\bibnamefont {Kubo}}, \bibinfo {author}
  {\bibfnamefont {M.}~\bibnamefont {Stern}}, \bibinfo {author} {\bibfnamefont
  {X.}~\bibnamefont {Zhou}}, \bibinfo {author} {\bibfnamefont {C.-C.}\
  \bibnamefont {Lo}}, \bibinfo {author} {\bibfnamefont {C.}~\bibnamefont
  {Weis}}, \bibinfo {author} {\bibfnamefont {T.}~\bibnamefont {Schenkel}},
  \bibinfo {author} {\bibfnamefont {M.}~\bibnamefont {Thewalt}}, \bibinfo
  {author} {\bibfnamefont {D.}~\bibnamefont {Vion}}, \bibinfo {author}
  {\bibfnamefont {D.}~\bibnamefont {Esteve}}, \bibinfo {author} {\bibfnamefont
  {B.}~\bibnamefont {Julsgaard}}, \bibinfo {author} {\bibfnamefont
  {K.}~\bibnamefont {Moelmer}}, \bibinfo {author} {\bibfnamefont
  {J.}~\bibnamefont {Morton}}, \ and\ \bibinfo {author} {\bibfnamefont
  {P.}~\bibnamefont {Bertet}},\ }\href {\doibase 10.1038/nnano.2015.282}
  {\bibfield  {journal} {\bibinfo  {journal} {Nature Nanotechnology}\ }\textbf
  {\bibinfo {volume} {11}},\ \bibinfo {pages} {253 } (\bibinfo {year}
  {2015})}\BibitemShut {NoStop}%
\bibitem [{\citenamefont {Eichler}\ \emph {et~al.}(2017)\citenamefont
  {Eichler}, \citenamefont {Sigillito}, \citenamefont {Lyon},\ and\
  \citenamefont {Petta}}]{eichler_electron_2017}%
  \BibitemOpen
  \bibfield  {author} {\bibinfo {author} {\bibfnamefont {C.}~\bibnamefont
  {Eichler}}, \bibinfo {author} {\bibfnamefont {A.~J.}\ \bibnamefont
  {Sigillito}}, \bibinfo {author} {\bibfnamefont {S.~A.}\ \bibnamefont {Lyon}},
  \ and\ \bibinfo {author} {\bibfnamefont {J.~R.}\ \bibnamefont {Petta}},\
  }\href {\doibase 10.1103/PhysRevLett.118.037701} {\bibfield  {journal}
  {\bibinfo  {journal} {Phys. Rev. Lett.}\ }\textbf {\bibinfo {volume} {118}},\
  \bibinfo {pages} {037701} (\bibinfo {year} {2017})}\BibitemShut {NoStop}%
\bibitem [{\citenamefont {Purcell}(1946)}]{purcell_spontaneous_1946}%
  \BibitemOpen
  \bibfield  {author} {\bibinfo {author} {\bibfnamefont {E.~M.}\ \bibnamefont
  {Purcell}},\ }\href@noop {} {\bibfield  {journal} {\bibinfo  {journal} {Phys.
  Rev.}\ }\textbf {\bibinfo {volume} {69}},\ \bibinfo {pages} {681} (\bibinfo
  {year} {1946})}\BibitemShut {NoStop}%
\bibitem [{\citenamefont {Goy}\ \emph {et~al.}(1983)\citenamefont {Goy},
  \citenamefont {Raimond}, \citenamefont {Gross},\ and\ \citenamefont
  {Haroche}}]{goy_observation_1983}%
  \BibitemOpen
  \bibfield  {author} {\bibinfo {author} {\bibfnamefont {P.}~\bibnamefont
  {Goy}}, \bibinfo {author} {\bibfnamefont {J.~M.}\ \bibnamefont {Raimond}},
  \bibinfo {author} {\bibfnamefont {M.}~\bibnamefont {Gross}}, \ and\ \bibinfo
  {author} {\bibfnamefont {S.}~\bibnamefont {Haroche}},\ }\href {\doibase
  10.1103/PhysRevLett.50.1903} {\bibfield  {journal} {\bibinfo  {journal}
  {Physical Review Letters}\ }\textbf {\bibinfo {volume} {50}},\ \bibinfo
  {pages} {1903} (\bibinfo {year} {1983})}\BibitemShut {NoStop}%
\bibitem [{\citenamefont {Bienfait}\ \emph {et~al.}(2016)\citenamefont
  {Bienfait}, \citenamefont {Pla}, \citenamefont {Kubo}, \citenamefont {Zhou},
  \citenamefont {Stern}, \citenamefont {Lo}, \citenamefont {Weis},
  \citenamefont {Schenkel}, \citenamefont {Vion}, \citenamefont {Esteve},
  \citenamefont {Morton},\ and\ \citenamefont
  {Bertet}}]{bienfait_controlling_2016}%
  \BibitemOpen
  \bibfield  {author} {\bibinfo {author} {\bibfnamefont {A.}~\bibnamefont
  {Bienfait}}, \bibinfo {author} {\bibfnamefont {J.}~\bibnamefont {Pla}},
  \bibinfo {author} {\bibfnamefont {Y.}~\bibnamefont {Kubo}}, \bibinfo {author}
  {\bibfnamefont {X.}~\bibnamefont {Zhou}}, \bibinfo {author} {\bibfnamefont
  {M.}~\bibnamefont {Stern}}, \bibinfo {author} {\bibfnamefont {C.-C.}\
  \bibnamefont {Lo}}, \bibinfo {author} {\bibfnamefont {C.}~\bibnamefont
  {Weis}}, \bibinfo {author} {\bibfnamefont {T.}~\bibnamefont {Schenkel}},
  \bibinfo {author} {\bibfnamefont {D.}~\bibnamefont {Vion}}, \bibinfo {author}
  {\bibfnamefont {D.}~\bibnamefont {Esteve}}, \bibinfo {author} {\bibfnamefont
  {J.}~\bibnamefont {Morton}}, \ and\ \bibinfo {author} {\bibfnamefont
  {P.}~\bibnamefont {Bertet}},\ }\href {\doibase doi:10.1038/nature16944}
  {\bibfield  {journal} {\bibinfo  {journal} {Nature}\ }\textbf {\bibinfo
  {volume} {531}},\ \bibinfo {pages} {74 } (\bibinfo {year}
  {2016})}\BibitemShut {NoStop}%
\bibitem [{\citenamefont {Probst}\ \emph {et~al.}(2017)\citenamefont {Probst},
  \citenamefont {Bienfait}, \citenamefont {Campagne-Ibarcq}, \citenamefont
  {Pla}, \citenamefont {Albanese}, \citenamefont {Barbosa}, \citenamefont
  {Schenkel}, \citenamefont {Vion}, \citenamefont {Esteve}, \citenamefont
  {Moelmer}, \citenamefont {Morton}, \citenamefont {Heeres},\ and\
  \citenamefont {Bertet}}]{probst_inductive-detection_2017}%
  \BibitemOpen
  \bibfield  {author} {\bibinfo {author} {\bibfnamefont {S.}~\bibnamefont
  {Probst}}, \bibinfo {author} {\bibfnamefont {A.}~\bibnamefont {Bienfait}},
  \bibinfo {author} {\bibfnamefont {P.}~\bibnamefont {Campagne-Ibarcq}},
  \bibinfo {author} {\bibfnamefont {J.~J.}\ \bibnamefont {Pla}}, \bibinfo
  {author} {\bibfnamefont {B.}~\bibnamefont {Albanese}}, \bibinfo {author}
  {\bibfnamefont {J.~F. D.~S.}\ \bibnamefont {Barbosa}}, \bibinfo {author}
  {\bibfnamefont {T.}~\bibnamefont {Schenkel}}, \bibinfo {author}
  {\bibfnamefont {D.}~\bibnamefont {Vion}}, \bibinfo {author} {\bibfnamefont
  {D.}~\bibnamefont {Esteve}}, \bibinfo {author} {\bibfnamefont
  {K.}~\bibnamefont {Moelmer}}, \bibinfo {author} {\bibfnamefont {J.~J.~L.}\
  \bibnamefont {Morton}}, \bibinfo {author} {\bibfnamefont {R.}~\bibnamefont
  {Heeres}}, \ and\ \bibinfo {author} {\bibfnamefont {P.}~\bibnamefont
  {Bertet}},\ }\href {\doibase 10.1063/1.5002540} {\bibfield  {journal}
  {\bibinfo  {journal} {Applied Physics Letters}\ }\textbf {\bibinfo {volume}
  {111}},\ \bibinfo {pages} {202604} (\bibinfo {year} {2017})}\BibitemShut
  {NoStop}%
\bibitem [{Note1()}]{Note1}%
  \BibitemOpen
  \bibinfo {note} {CST Microwave Studios$\protect \textsuperscript {\relax
  \protect \fontsize {5}{6}\protect \selectfont \textregistered }$}\BibitemShut
  {NoStop}%
\bibitem [{\citenamefont {Feher}(1959)}]{feher_electron_1959}%
  \BibitemOpen
  \bibfield  {author} {\bibinfo {author} {\bibfnamefont {G.}~\bibnamefont
  {Feher}},\ }\href {\doibase 10.1103/PhysRev.114.1219} {\bibfield  {journal}
  {\bibinfo  {journal} {Phys. Rev.}\ }\textbf {\bibinfo {volume} {114}},\
  \bibinfo {pages} {1219} (\bibinfo {year} {1959})}\BibitemShut {NoStop}%
\bibitem [{\citenamefont {Morley}\ \emph {et~al.}(2010)\citenamefont {Morley},
  \citenamefont {Warner}, \citenamefont {Stoneham}, \citenamefont {Greenland},
  \citenamefont {van Tol}, \citenamefont {Kay},\ and\ \citenamefont
  {Aeppli}}]{morley_initialization_2010}%
  \BibitemOpen
  \bibfield  {author} {\bibinfo {author} {\bibfnamefont {G.~W.}\ \bibnamefont
  {Morley}}, \bibinfo {author} {\bibfnamefont {M.}~\bibnamefont {Warner}},
  \bibinfo {author} {\bibfnamefont {A.~M.}\ \bibnamefont {Stoneham}}, \bibinfo
  {author} {\bibfnamefont {P.~T.}\ \bibnamefont {Greenland}}, \bibinfo {author}
  {\bibfnamefont {J.}~\bibnamefont {van Tol}}, \bibinfo {author} {\bibfnamefont
  {C.~W.}\ \bibnamefont {Kay}}, \ and\ \bibinfo {author} {\bibfnamefont
  {G.}~\bibnamefont {Aeppli}},\ }\href {\doibase doi:10.1038/nmat2828}
  {\bibfield  {journal} {\bibinfo  {journal} {Nature materials}\ }\textbf
  {\bibinfo {volume} {9}},\ \bibinfo {pages} {725} (\bibinfo {year}
  {2010})}\BibitemShut {NoStop}%
\bibitem [{\citenamefont {Mohammady}, \citenamefont {Morley},\ and\
  \citenamefont {Monteiro}(2010)}]{mohammady_bismuth_2010}%
  \BibitemOpen
  \bibfield  {author} {\bibinfo {author} {\bibfnamefont {M.~H.}\ \bibnamefont
  {Mohammady}}, \bibinfo {author} {\bibfnamefont {G.~W.}\ \bibnamefont
  {Morley}}, \ and\ \bibinfo {author} {\bibfnamefont {T.~S.}\ \bibnamefont
  {Monteiro}},\ }\href {\doibase 10.1103/physrevlett.105.067602} {\bibfield
  {journal} {\bibinfo  {journal} {Physical Review Letters}\ }\textbf {\bibinfo
  {volume} {105}} (\bibinfo {year} {2010}),\
  10.1103/physrevlett.105.067602}\BibitemShut {NoStop}%
\bibitem [{\citenamefont {George}\ \emph {et~al.}(2010)\citenamefont {George},
  \citenamefont {Witzel}, \citenamefont {Riemann}, \citenamefont {Abrosimov},
  \citenamefont {Nötzel}, \citenamefont {Thewalt},\ and\ \citenamefont
  {Morton}}]{george_electron_2010}%
  \BibitemOpen
  \bibfield  {author} {\bibinfo {author} {\bibfnamefont {R.~E.}\ \bibnamefont
  {George}}, \bibinfo {author} {\bibfnamefont {W.}~\bibnamefont {Witzel}},
  \bibinfo {author} {\bibfnamefont {H.}~\bibnamefont {Riemann}}, \bibinfo
  {author} {\bibfnamefont {N.~V.}\ \bibnamefont {Abrosimov}}, \bibinfo {author}
  {\bibfnamefont {N.}~\bibnamefont {Nötzel}}, \bibinfo {author} {\bibfnamefont
  {M.~L.~W.}\ \bibnamefont {Thewalt}}, \ and\ \bibinfo {author} {\bibfnamefont
  {J.~J.~L.}\ \bibnamefont {Morton}},\ }\href {\doibase
  10.1103/PhysRevLett.105.067601} {\bibfield  {journal} {\bibinfo  {journal}
  {Physical Review Letters}\ }\textbf {\bibinfo {volume} {105}},\ \bibinfo
  {pages} {067601} (\bibinfo {year} {2010})}\BibitemShut {NoStop}%
\bibitem [{\citenamefont {Wolfowicz}\ \emph {et~al.}(2013)\citenamefont
  {Wolfowicz}, \citenamefont {Tyryshkin}, \citenamefont {George}, \citenamefont
  {Riemann}, \citenamefont {Abrosimov}, \citenamefont {Becker}, \citenamefont
  {Pohl}, \citenamefont {Thewalt}, \citenamefont {Lyon},\ and\ \citenamefont
  {Morton}}]{wolfowicz_atomic_2013}%
  \BibitemOpen
  \bibfield  {author} {\bibinfo {author} {\bibfnamefont {G.}~\bibnamefont
  {Wolfowicz}}, \bibinfo {author} {\bibfnamefont {A.~M.}\ \bibnamefont
  {Tyryshkin}}, \bibinfo {author} {\bibfnamefont {R.~E.}\ \bibnamefont
  {George}}, \bibinfo {author} {\bibfnamefont {H.}~\bibnamefont {Riemann}},
  \bibinfo {author} {\bibfnamefont {N.~V.}\ \bibnamefont {Abrosimov}}, \bibinfo
  {author} {\bibfnamefont {P.}~\bibnamefont {Becker}}, \bibinfo {author}
  {\bibfnamefont {H.-J.}\ \bibnamefont {Pohl}}, \bibinfo {author}
  {\bibfnamefont {M.~L.~W.}\ \bibnamefont {Thewalt}}, \bibinfo {author}
  {\bibfnamefont {S.~a.}\ \bibnamefont {Lyon}}, \ and\ \bibinfo {author}
  {\bibfnamefont {J.~J.~L.}\ \bibnamefont {Morton}},\ }\href {\doibase
  10.1038/nnano.2013.117} {\bibfield  {journal} {\bibinfo  {journal} {Nature
  Nanotechnology}\ }\textbf {\bibinfo {volume} {8}},\ \bibinfo {pages} {561}
  (\bibinfo {year} {2013})}\BibitemShut {NoStop}%
\bibitem [{\citenamefont {Zhou}\ \emph {et~al.}(2014)\citenamefont {Zhou},
  \citenamefont {Schmitt}, \citenamefont {Bertet}, \citenamefont {Vion},
  \citenamefont {Wustmann}, \citenamefont {Shumeiko},\ and\ \citenamefont
  {Esteve}}]{zhou_high-gain_2014}%
  \BibitemOpen
  \bibfield  {author} {\bibinfo {author} {\bibfnamefont {X.}~\bibnamefont
  {Zhou}}, \bibinfo {author} {\bibfnamefont {V.}~\bibnamefont {Schmitt}},
  \bibinfo {author} {\bibfnamefont {P.}~\bibnamefont {Bertet}}, \bibinfo
  {author} {\bibfnamefont {D.}~\bibnamefont {Vion}}, \bibinfo {author}
  {\bibfnamefont {W.}~\bibnamefont {Wustmann}}, \bibinfo {author}
  {\bibfnamefont {V.}~\bibnamefont {Shumeiko}}, \ and\ \bibinfo {author}
  {\bibfnamefont {D.}~\bibnamefont {Esteve}},\ }\href {\doibase
  10.1103/PhysRevB.89.214517} {\bibfield  {journal} {\bibinfo  {journal} {Phys.
  Rev. B}\ }\textbf {\bibinfo {volume} {89}},\ \bibinfo {pages} {214517}
  (\bibinfo {year} {2014})}\BibitemShut {NoStop}%
\bibitem [{\citenamefont {Martinis}\ \emph {et~al.}(2005)\citenamefont
  {Martinis}, \citenamefont {Cooper}, \citenamefont {McDermott}, \citenamefont
  {Steffen}, \citenamefont {Ansmann}, \citenamefont {Osborn}, \citenamefont
  {Cicak}, \citenamefont {Oh}, \citenamefont {Pappas}, \citenamefont
  {Simmonds},\ and\ \citenamefont {Yu}}]{martinis_decoherence_2005}%
  \BibitemOpen
  \bibfield  {author} {\bibinfo {author} {\bibfnamefont {J.~M.}\ \bibnamefont
  {Martinis}}, \bibinfo {author} {\bibfnamefont {K.~B.}\ \bibnamefont
  {Cooper}}, \bibinfo {author} {\bibfnamefont {R.}~\bibnamefont {McDermott}},
  \bibinfo {author} {\bibfnamefont {M.}~\bibnamefont {Steffen}}, \bibinfo
  {author} {\bibfnamefont {M.}~\bibnamefont {Ansmann}}, \bibinfo {author}
  {\bibfnamefont {K.~D.}\ \bibnamefont {Osborn}}, \bibinfo {author}
  {\bibfnamefont {K.}~\bibnamefont {Cicak}}, \bibinfo {author} {\bibfnamefont
  {S.}~\bibnamefont {Oh}}, \bibinfo {author} {\bibfnamefont {D.~P.}\
  \bibnamefont {Pappas}}, \bibinfo {author} {\bibfnamefont {R.~W.}\
  \bibnamefont {Simmonds}}, \ and\ \bibinfo {author} {\bibfnamefont {C.~C.}\
  \bibnamefont {Yu}},\ }\href {\doibase 10.1103/PhysRevLett.95.210503}
  {\bibfield  {journal} {\bibinfo  {journal} {Physical Review Letters}\
  }\textbf {\bibinfo {volume} {95}},\ \bibinfo {pages} {210503} (\bibinfo
  {year} {2005})}\BibitemShut {NoStop}%
\bibitem [{\citenamefont {Levenson-Falk}, \citenamefont {Vijay},\ and\
  \citenamefont {Siddiqi}(2011)}]{levenson-falk_nonlinear_2011}%
  \BibitemOpen
  \bibfield  {author} {\bibinfo {author} {\bibfnamefont {E.~M.}\ \bibnamefont
  {Levenson-Falk}}, \bibinfo {author} {\bibfnamefont {R.}~\bibnamefont
  {Vijay}}, \ and\ \bibinfo {author} {\bibfnamefont {I.}~\bibnamefont
  {Siddiqi}},\ }\href {\doibase 10.1063/1.3570693} {\bibfield  {journal}
  {\bibinfo  {journal} {Applied Physics Letters}\ }\textbf {\bibinfo {volume}
  {98}},\ \bibinfo {pages} {123115} (\bibinfo {year} {2011})}\BibitemShut
  {NoStop}%
\bibitem [{\citenamefont {Asfaw}\ \emph {et~al.}(2017)\citenamefont {Asfaw},
  \citenamefont {Sigillito}, \citenamefont {Tyryshkin}, \citenamefont
  {Schenkel},\ and\ \citenamefont {Lyon}}]{asfaw_multi-frequency_2017}%
  \BibitemOpen
  \bibfield  {author} {\bibinfo {author} {\bibfnamefont {A.~T.}\ \bibnamefont
  {Asfaw}}, \bibinfo {author} {\bibfnamefont {A.~J.}\ \bibnamefont
  {Sigillito}}, \bibinfo {author} {\bibfnamefont {A.~M.}\ \bibnamefont
  {Tyryshkin}}, \bibinfo {author} {\bibfnamefont {T.}~\bibnamefont {Schenkel}},
  \ and\ \bibinfo {author} {\bibfnamefont {S.~A.}\ \bibnamefont {Lyon}},\
  }\href {\doibase 10.1063/1.4993930} {\bibfield  {journal} {\bibinfo
  {journal} {Applied Physics Letters}\ }\textbf {\bibinfo {volume} {111}},\
  \bibinfo {pages} {032601} (\bibinfo {year} {2017})}\BibitemShut {NoStop}%
\bibitem [{\citenamefont {Ranjan}\ \emph {et~al.}(2020)\citenamefont {Ranjan},
  \citenamefont {Probst}, \citenamefont {Albanese}, \citenamefont {Doll},
  \citenamefont {Jacquot}, \citenamefont {Flurin}, \citenamefont {Heeres},
  \citenamefont {Vion}, \citenamefont {Esteve}, \citenamefont {Morton},\ and\
  \citenamefont {Bertet}}]{ranjan_pulsed_2020}%
  \BibitemOpen
  \bibfield  {author} {\bibinfo {author} {\bibfnamefont {V.}~\bibnamefont
  {Ranjan}}, \bibinfo {author} {\bibfnamefont {S.}~\bibnamefont {Probst}},
  \bibinfo {author} {\bibfnamefont {B.}~\bibnamefont {Albanese}}, \bibinfo
  {author} {\bibfnamefont {A.}~\bibnamefont {Doll}}, \bibinfo {author}
  {\bibfnamefont {O.}~\bibnamefont {Jacquot}}, \bibinfo {author} {\bibfnamefont
  {E.}~\bibnamefont {Flurin}}, \bibinfo {author} {\bibfnamefont
  {R.}~\bibnamefont {Heeres}}, \bibinfo {author} {\bibfnamefont
  {D.}~\bibnamefont {Vion}}, \bibinfo {author} {\bibfnamefont {D.}~\bibnamefont
  {Esteve}}, \bibinfo {author} {\bibfnamefont {J.~J.~L.}\ \bibnamefont
  {Morton}}, \ and\ \bibinfo {author} {\bibfnamefont {P.}~\bibnamefont
  {Bertet}},\ }\href {\doibase https://doi.org/10.1016/j.jmr.2019.106662}
  {\bibfield  {journal} {\bibinfo  {journal} {Journal of Magnetic Resonance}\
  }\textbf {\bibinfo {volume} {310}},\ \bibinfo {pages} {106662} (\bibinfo
  {year} {2020})}\BibitemShut {NoStop}%
\bibitem [{\citenamefont {Mansir}\ \emph {et~al.}(2018)\citenamefont {Mansir},
  \citenamefont {Conti}, \citenamefont {Zeng}, \citenamefont {Pla},
  \citenamefont {Bertet}, \citenamefont {Swift}, \citenamefont {Van~de Walle},
  \citenamefont {Thewalt}, \citenamefont {Sklenard}, \citenamefont {Niquet},\
  and\ \citenamefont {Morton}}]{mansir_linear_2018}%
  \BibitemOpen
  \bibfield  {author} {\bibinfo {author} {\bibfnamefont {J.}~\bibnamefont
  {Mansir}}, \bibinfo {author} {\bibfnamefont {P.}~\bibnamefont {Conti}},
  \bibinfo {author} {\bibfnamefont {Z.}~\bibnamefont {Zeng}}, \bibinfo {author}
  {\bibfnamefont {J.}~\bibnamefont {Pla}}, \bibinfo {author} {\bibfnamefont
  {P.}~\bibnamefont {Bertet}}, \bibinfo {author} {\bibfnamefont
  {M.}~\bibnamefont {Swift}}, \bibinfo {author} {\bibfnamefont
  {C.}~\bibnamefont {Van~de Walle}}, \bibinfo {author} {\bibfnamefont
  {M.}~\bibnamefont {Thewalt}}, \bibinfo {author} {\bibfnamefont
  {B.}~\bibnamefont {Sklenard}}, \bibinfo {author} {\bibfnamefont
  {Y.}~\bibnamefont {Niquet}}, \ and\ \bibinfo {author} {\bibfnamefont
  {J.}~\bibnamefont {Morton}},\ }\href {\doibase
  10.1103/PhysRevLett.120.167701} {\bibfield  {journal} {\bibinfo  {journal}
  {Physical Review Letters}\ }\textbf {\bibinfo {volume} {120}},\ \bibinfo
  {pages} {167701} (\bibinfo {year} {2018})}\BibitemShut {NoStop}%
\bibitem [{\citenamefont {Pla}\ \emph {et~al.}(2018)\citenamefont {Pla},
  \citenamefont {Bienfait}, \citenamefont {Pica}, \citenamefont {Mansir},
  \citenamefont {Mohiyaddin}, \citenamefont {Zeng}, \citenamefont {Niquet},
  \citenamefont {Morello}, \citenamefont {Schenkel}, \citenamefont {Morton},\
  and\ \citenamefont {Bertet}}]{pla_strain-induced_2018}%
  \BibitemOpen
  \bibfield  {author} {\bibinfo {author} {\bibfnamefont {J.}~\bibnamefont
  {Pla}}, \bibinfo {author} {\bibfnamefont {A.}~\bibnamefont {Bienfait}},
  \bibinfo {author} {\bibfnamefont {G.}~\bibnamefont {Pica}}, \bibinfo {author}
  {\bibfnamefont {J.}~\bibnamefont {Mansir}}, \bibinfo {author} {\bibfnamefont
  {F.}~\bibnamefont {Mohiyaddin}}, \bibinfo {author} {\bibfnamefont
  {Z.}~\bibnamefont {Zeng}}, \bibinfo {author} {\bibfnamefont {Y.}~\bibnamefont
  {Niquet}}, \bibinfo {author} {\bibfnamefont {A.}~\bibnamefont {Morello}},
  \bibinfo {author} {\bibfnamefont {T.}~\bibnamefont {Schenkel}}, \bibinfo
  {author} {\bibfnamefont {J.}~\bibnamefont {Morton}}, \ and\ \bibinfo {author}
  {\bibfnamefont {P.}~\bibnamefont {Bertet}},\ }\href {\doibase
  10.1103/PhysRevApplied.9.044014} {\bibfield  {journal} {\bibinfo  {journal}
  {Physical Review Applied}\ }\textbf {\bibinfo {volume} {9}},\ \bibinfo
  {pages} {044014} (\bibinfo {year} {2018})}\BibitemShut {NoStop}%
\bibitem [{\citenamefont {Probst}\ \emph {et~al.}(2020)\citenamefont {Probst},
  \citenamefont {Zhang}, \citenamefont {Rancic}, \citenamefont {Ranjan},
  \citenamefont {Dantec}, \citenamefont {Zhong}, \citenamefont {Albanese},
  \citenamefont {Doll}, \citenamefont {Liu}, \citenamefont {Morton},
  \citenamefont {Chanelilere}, \citenamefont {Goldner}, \citenamefont {Vion},
  \citenamefont {Esteve},\ and\ \citenamefont
  {Bertet}}]{probst_hyperfine_2020}%
  \BibitemOpen
  \bibfield  {author} {\bibinfo {author} {\bibfnamefont {S.}~\bibnamefont
  {Probst}}, \bibinfo {author} {\bibfnamefont {G.~L.}\ \bibnamefont {Zhang}},
  \bibinfo {author} {\bibfnamefont {M.}~\bibnamefont {Rancic}}, \bibinfo
  {author} {\bibfnamefont {V.}~\bibnamefont {Ranjan}}, \bibinfo {author}
  {\bibfnamefont {M.~L.}\ \bibnamefont {Dantec}}, \bibinfo {author}
  {\bibfnamefont {Z.}~\bibnamefont {Zhong}}, \bibinfo {author} {\bibfnamefont
  {B.}~\bibnamefont {Albanese}}, \bibinfo {author} {\bibfnamefont
  {A.}~\bibnamefont {Doll}}, \bibinfo {author} {\bibfnamefont {R.~B.}\
  \bibnamefont {Liu}}, \bibinfo {author} {\bibfnamefont {J.~J.~L.}\
  \bibnamefont {Morton}}, \bibinfo {author} {\bibfnamefont {T.}~\bibnamefont
  {Chanelilere}}, \bibinfo {author} {\bibfnamefont {P.}~\bibnamefont
  {Goldner}}, \bibinfo {author} {\bibfnamefont {D.}~\bibnamefont {Vion}},
  \bibinfo {author} {\bibfnamefont {D.}~\bibnamefont {Esteve}}, \ and\ \bibinfo
  {author} {\bibfnamefont {P.}~\bibnamefont {Bertet}},\ }\href
  {http://arxiv.org/abs/2001.04854} {\bibfield  {journal} {\bibinfo  {journal}
  {arXiv:2001.04854 [quant-ph]}\ } (\bibinfo {year} {2020})},\ \bibinfo {note}
  {arXiv: 2001.04854}\BibitemShut {NoStop}%
\bibitem [{\citenamefont {Zhu}\ \emph {et~al.}(2011)\citenamefont {Zhu},
  \citenamefont {Saito}, \citenamefont {Kemp}, \citenamefont {Kakuyanagi},
  \citenamefont {Karimoto}, \citenamefont {Nakano}, \citenamefont {Munro},
  \citenamefont {Tokura}, \citenamefont {Everitt}, \citenamefont {Nemoto},
  \citenamefont {Kasu}, \citenamefont {Mizuochi},\ and\ \citenamefont
  {Semba}}]{zhu_coherent_2011}%
  \BibitemOpen
  \bibfield  {author} {\bibinfo {author} {\bibfnamefont {X.}~\bibnamefont
  {Zhu}}, \bibinfo {author} {\bibfnamefont {S.}~\bibnamefont {Saito}}, \bibinfo
  {author} {\bibfnamefont {A.}~\bibnamefont {Kemp}}, \bibinfo {author}
  {\bibfnamefont {K.}~\bibnamefont {Kakuyanagi}}, \bibinfo {author}
  {\bibfnamefont {S.-i.}\ \bibnamefont {Karimoto}}, \bibinfo {author}
  {\bibfnamefont {H.}~\bibnamefont {Nakano}}, \bibinfo {author} {\bibfnamefont
  {W.~J.}\ \bibnamefont {Munro}}, \bibinfo {author} {\bibfnamefont
  {Y.}~\bibnamefont {Tokura}}, \bibinfo {author} {\bibfnamefont {M.~S.}\
  \bibnamefont {Everitt}}, \bibinfo {author} {\bibfnamefont {K.}~\bibnamefont
  {Nemoto}}, \bibinfo {author} {\bibfnamefont {M.}~\bibnamefont {Kasu}},
  \bibinfo {author} {\bibfnamefont {N.}~\bibnamefont {Mizuochi}}, \ and\
  \bibinfo {author} {\bibfnamefont {K.}~\bibnamefont {Semba}},\ }\href
  {\doibase 10.1038/nature10462} {\bibfield  {journal} {\bibinfo  {journal}
  {Nature}\ }\textbf {\bibinfo {volume} {478}},\ \bibinfo {pages} {221}
  (\bibinfo {year} {2011})}\BibitemShut {NoStop}%
\bibitem [{\citenamefont {Bienfait}\ \emph {et~al.}(2017)\citenamefont
  {Bienfait}, \citenamefont {Campagne-Ibarcq}, \citenamefont {Kiilerich},
  \citenamefont {Zhou}, \citenamefont {Probst}, \citenamefont {Pla},
  \citenamefont {Schenkel}, \citenamefont {Vion}, \citenamefont {Esteve},
  \citenamefont {Morton}, \citenamefont {Moelmer},\ and\ \citenamefont
  {Bertet}}]{bienfait_magnetic_2017}%
  \BibitemOpen
  \bibfield  {author} {\bibinfo {author} {\bibfnamefont {A.}~\bibnamefont
  {Bienfait}}, \bibinfo {author} {\bibfnamefont {P.}~\bibnamefont
  {Campagne-Ibarcq}}, \bibinfo {author} {\bibfnamefont {A.}~\bibnamefont
  {Kiilerich}}, \bibinfo {author} {\bibfnamefont {X.}~\bibnamefont {Zhou}},
  \bibinfo {author} {\bibfnamefont {S.}~\bibnamefont {Probst}}, \bibinfo
  {author} {\bibfnamefont {J.}~\bibnamefont {Pla}}, \bibinfo {author}
  {\bibfnamefont {T.}~\bibnamefont {Schenkel}}, \bibinfo {author}
  {\bibfnamefont {D.}~\bibnamefont {Vion}}, \bibinfo {author} {\bibfnamefont
  {D.}~\bibnamefont {Esteve}}, \bibinfo {author} {\bibfnamefont
  {J.}~\bibnamefont {Morton}}, \bibinfo {author} {\bibfnamefont
  {K.}~\bibnamefont {Moelmer}}, \ and\ \bibinfo {author} {\bibfnamefont
  {P.}~\bibnamefont {Bertet}},\ }\href {\doibase 10.1103/PhysRevX.7.041011}
  {\bibfield  {journal} {\bibinfo  {journal} {Physical Review X}\ }\textbf
  {\bibinfo {volume} {7}},\ \bibinfo {pages} {041011} (\bibinfo {year}
  {2017})}\BibitemShut {NoStop}%
\bibitem [{\citenamefont {Mentink-Vigier}\ \emph {et~al.}(2013)\citenamefont
  {Mentink-Vigier}, \citenamefont {Collauto}, \citenamefont {Feintuch},
  \citenamefont {Kaminker}, \citenamefont {Tarle},\ and\ \citenamefont
  {Goldfarb}}]{mentink-vigier_increasing_2013}%
  \BibitemOpen
  \bibfield  {author} {\bibinfo {author} {\bibfnamefont {F.}~\bibnamefont
  {Mentink-Vigier}}, \bibinfo {author} {\bibfnamefont {A.}~\bibnamefont
  {Collauto}}, \bibinfo {author} {\bibfnamefont {A.}~\bibnamefont {Feintuch}},
  \bibinfo {author} {\bibfnamefont {I.}~\bibnamefont {Kaminker}}, \bibinfo
  {author} {\bibfnamefont {V.}~\bibnamefont {Tarle}}, \ and\ \bibinfo {author}
  {\bibfnamefont {D.}~\bibnamefont {Goldfarb}},\ }\href {\doibase
  10.1016/j.jmr.2013.08.012} {\bibfield  {journal} {\bibinfo  {journal}
  {Journal of Magnetic Resonance}\ }\textbf {\bibinfo {volume} {236}},\
  \bibinfo {pages} {117 } (\bibinfo {year} {2013})}\BibitemShut {NoStop}%
\bibitem [{\citenamefont {Geim}\ and\ \citenamefont
  {Grigorieva}(2013)}]{geim_van_2013}%
  \BibitemOpen
  \bibfield  {author} {\bibinfo {author} {\bibfnamefont {A.~K.}\ \bibnamefont
  {Geim}}\ and\ \bibinfo {author} {\bibfnamefont {I.~V.}\ \bibnamefont
  {Grigorieva}},\ }\href {\doibase 10.1038/nature12385} {\bibfield  {journal}
  {\bibinfo  {journal} {Nature}\ }\textbf {\bibinfo {volume} {499}},\ \bibinfo
  {pages} {419} (\bibinfo {year} {2013})}\BibitemShut {NoStop}%
\end{thebibliography}

\end{document}